# DeLottery: A Novel Decentralized Lottery System Based on Blockchain Technology

Zhifeng Jia, Rui Chen
UM-SJTU Joint Institute
Shanghai Jiao Tong University, Shanghai, China
{fergusjia, 2016jichenru}@sjtu.edu.cn

Jie Li
Department of Computer Science
Shanghai Jiao Tong University, Shanghai, China
lijiecs@sjtu.edu.cn

## ABSTRACT
In this paper, we design DeLottery, a decentralized lottery and gambling system based on block chain technology and smart contracts. Lottery is a classical form of entertainment and charity for centuries. Facing the bottleneck of the combination between lottery and information technology, we use smart contracts and blockchain in decentralized, intelligent, and secure systems for lottery industries. Moreover, we are inspired by the algorithm of RANDAO, an outstanding way of random number generation in blockchain scenario. The components and the functions of the novel system are described in details. We implement DeLottery in a blockchain network and show functioning procedure and security of the proposed lottery system.

## CCS Concepts

• **Applied computing ~ Secure online transactions**; Electronic funds transfer

• Information systems ~ Secure online transactions

## Keywords
lottery system, decentralized, blockchain, smart contract

## 1. INTRODUCTION

Lottery is a financial activity where people pay money for their bets, in most cases, certain sequences of numbers, and then they have chances to win prizes [1]. The current prevailing lottery system functions as follows:

(1) A lottery company starts a lottery;

(2) People make bets and pay money;

(3) The lottery company generate winner numbers randomly;

(4) Winners collect their prizes.

Suppose that there are certain number participants taken part in a lottery event, hosted by a lottery company or organization. For a typical event, the parameters are listed in Table 1. The fairness of lotteries depends mostly on the third parties [12].

These companies as the centers of these activities are trusted by participants. They set the rules and regulations which are supposed to be equal to every individual. The lottery events are usually hosted by lottery companies or governments, and the whole procedure of a lottery event often takes an entire day. The clumsiness of a host company or organization is preventing the traditional lottery system from high efficiency.

When it comes to large traditional lottery events, it often takes the third party around 24 hours to collect the lottery tickets, making the traditional lottery process quite time-consuming and inconvenient for individuals to host lottery events. People have been aware of the inconvenience, and stepped forward with the proposal of web-based lottery system.

Back in 1999, David Leason and Scott L. Sullivan developed a type of online lottery system with centralized feature. The system design was among one of the greatest contributions to the traditional lottery system, and it fulfilled people's need for instantly hosting lottery events in a relatively convenient way. In the last decade, a large number of lottery companies emerge, and influenced the traditional lottery business. However, the covenant-lite online lottery systems often fail to guarantee their users with the fundamental need for security, both in the sense of information security that their personal information is not leaked, and property security that their legal rewards can be guaranteed.

**Table 1: Traditional Lottery System Parameters**

| Name | Type | Description |
|---|---|---|
| $n$ | $n \in \mathbb{N}$ | the number of participants involved in the lottery process |
| $C$ | nonempty set | the set of all possible guesses of the lottery (treated as whole set) |
| $f_i$ | $f_i \in \mathbb{R}, i \in \{1, ..., n\}$ | the original amount of money of participant $i$ |
| $g_i$ | $g_i \subset \mathbb{R}, i \in \{1, ..., n\}$ | the set of guesses made by participant $i$ |
| $G$ | $G \subset C$ | the set of possible values of the winning results |
| $W$ | $W \subset G$ | the set of winning results of this lottery event |
| $t$ | $t \in \mathbb{R}$ | transaction fee deducted by the lottery company host |
| $p$ | $p \in \mathbb{R}$ | the total value of the accumulated reward |
| $s$ | $p \in \mathbb{N}$ | price of each guess |

Plenty of illegal lottery companies are gradually banned by governments during the past 10 years, and online lottery system has become a sensitive issue.

Concluding from above, the current lottery systems, whether traditional or online, are centralized ones that have the following potential problems:

(1) The traditional lottery procedure takes a relatively long period, and it is quite inconvenient for individuals to hold a lottery event in a traditional way.

(2) If the third party fails to be totally fair while producing the winners, benefits of the players are hurt.

(3) It is possible that the third party pay the winners a less amount of money or even nothing so that the third party itself can have some more interests.

We will discuss in the following section about the unique characteristics that blockchain and smart contracts technologies possess, and how we use these properties to replace the functions of a third party in existing systems.

## 2. PRELIMINARIES

In this section, we explain the blockchain technology and smart contract which will be used in our proposed system.

### 2.1 Blockchain

Blockchain technology [4] is a peer-to-peer networking technology which allows secured data storage and operations. The technology is based on public-key cryptography, and each account on chain is provided with a private key and a public key. The private key is kept by the account encrypted, while the public key is shared among all other accounts. At the beginning of a transaction, the receiver of the digital currencies sends his public key to the sender of the money by a hash-based digital signature [17]. The public keys are also considered as the addresses stored in the blockchain, and each coin is associated with an address. The novelty lying in the transaction process is that it is "enabled without disclosing one's identity". Thus, the security of personal information can be better protected compared to the original online and offline trading systems. The features of blockchain are supported by the following technologies:

- E-signature: Every piece of data needs a E-signature from its provider's account, which is the private key of that account. This private key is generated by HASH256 function using its unique public key.

*private_key = HASH256(public_key)*

- This HASH256 function has the property that one can always check whether the public key and private key are a match but it is nearly impossible to derive the private key from the public key. After the new data is posted on the blockchain, every other user can verify this E-signature to see whether the data is legal.

- Blocks linking: The data is stored on "blocks". Once a block is used up, a new block is needed for storage. At the beginning of each block there is a sequence of numbers which is the hash function output of the previous block's sequence. The blocks are linked together as a chain.

### 2.2 Smart Contract

Smart contract [9] is "a computerized transaction protocol that executes the terms of a contract". It is a kind of program deployed on the addresses of the accounts that has these following features:

- The inside logic of a smart contract is usually like a finite state machine. It is event-triggered, meaning that certain input event or certain function calling can trigger the smart contract to do previously designed operations.

- Once the smart contract is deployed, it cannot be further modified. As long as the requirements are satisfied, corresponding operations would surely be executed.

- Usually, the smart contract is deployed based on the blockchain. This means that the smart contract itself can be viewed by all the nodes on the blockchain and so can its operations and its users' participation. For example in Ethereum, the smart contract can be coded in language Solidity [3].

- The structure of a smart contract written in Solidity is similar to that of a class in C++ or Java.

Given these features, many social activities can now be programmed into smart contracts so that they would become more secure and automated [4][9][5][8][16]. Usually the role of a third party is no longer needed for these activities and their efficiency is greatly increased. Learning from these designs, we created a prototype smart contract for a lottery system.

## 3. SECURITY CONCERNS AND RANDAO

Although the properties of blockchain technology determine that the lottery system based on blockchain and smart contracts are free from the fairness and efficiency concerns of current lottery system designs, security issues still exist in blockchain-based lottery systems. Emin Gün Sirer expressed the security concerns in his paper "How Not To Run A Blockchain Lottery", and listed the weaknesses of an arbitrary lottery system [13].

- Node attack: The combination of hash values of multiple blocks cannot prevent the last transaction node from completely controlling the result;

- Random number generation: It is difficult to generate fair and uncontrolled random numbers in blockchain;

- Unfairness: The possibility of each lottery share to win the prize is different, meaning that players can attend the lottery event at certain period of time to bet for higher winning rates.

- Sybil attack: Blockchain technology is vulnerable to Sybil attack.

### 3.1 Node Attack and Sybil Attack

Random number generation is essential in a lottery process, and in hash functions, the usual way to generate a random number is Keccak [2], which is selected to be the standard cryptographic hash function for SHA-3. However, this method is vulnerable to hostile attacks from nodes, as is demonstrated in 2012 [6]. When one user is calling a function in the smart contract deployed, it will be broadcast to the transaction nodes on the network. Transaction nodes gather a certain amount of works broadcast to them, and publish a block wrapping up its Proof of Work (PoW) [15]. Other transaction nodes would stop dealing with this PoW, and continue to deal with another one. This leaves potential security issues, as transaction nodes can include the random process into every PoW it is dealing with, without informing other transaction nodes. If the random number results in a prize or refund, then the transaction nodes may add the random number generation work into its PoW, thus leading the accounts at the transaction nodes to win the prize.

Otherwise, the node would hide it from the proof of work, thus avoiding the payment process. Algorithm 1 illustrates the process by focusing on the working function of an arbitrary transaction node on a blockchain installed with a lottery system which generates the prize results by Keccak (or other methods that are integrated in the smart contract). Assume that the proof of work contains $\{E_1, E_2, ..., E_n\}$, and the lottery event EL is the lottery event that has just generated the winning result set

$W = \{w_1, ..., w_k\}$.

---
**Algorithm 1** Principle of a Node attack
---
**Require:** Proof of Work ($PoW = \{E_1, E_2, ..., E_n\}$), Lottery Event $E_L$;
**Ensure:** Processed Proof of Work ($PoW_{new}$);
1: Make a guess with the guess result set $g$.
2: **if** $g \subset W$ **then**
3:     $PoW_{new} = E_L \bigcup PoW$;
4: **else**
5:     $PoW_{new} = PoW$;
6: **end if**
    **return** $PoW_{new}$;
---

On the other hand, Sybil attack functions in a more difficult way as shown in Algorithm 2. Effective solutions to Sybil attack have been found out, mainly through Pow and identity certification.

• PoW: Proof of Work requires the node to pass certain calculation capability test to prove that it is a real node on chain. This method increases the cost of a Sybil attack, especially when the number of fake nodes is large.

• Identity certification: For decentralized blockchain systems, the certification requires all valid nodes that have passed the certification to agree on authorizing the new node to enter the blockchain network.

For blockchain-based lottery systems, especially when the system is applied to a small scale network with participating devices relatively weak in computing power, the cost to launch a Sybil attack can be greatly increased. With less and less profit attackers can gain, they will give up on attacking the lottery system for profit.

---
**Algorithm 2** Principle of a Sybil attack
---
**Require:** A blockchain with $n$ addresses $B = \{p_1, ..., p_n\}$, an arbitrary Sybil address $p_s$;
1: $p_s$ generates $k$ fake public key values according to the public keys of target addresses ($k > \frac{n}{2}$);
2: Broadcast fake addresses to other nodes;
3: Redirect requests passing through $p_s$, or return false values to other nodes.;
---

## 3.2 RANDAO

In order to deal the node attack problem, we make use of RANDAO, a verifiable random number generation method [10]. One of the fundamental restrictions that RANDAO set up is that transaction node addresses are prohibited to use RANDAO service, so the node attack issue can be avoided. This algorithm is open source, and uses following steps to generate random numbers.

• Collecting effective sha3($s$)

All players willing to take part in the production should send m amount of ETH deposit to Contract C within the specific window period, along with the sha3($s$) of a randomly selected number $s$.

• Collecting effective $s$

After the first step and within the specific window period of the second step, all producers that successfully submitted their sha3($s$) should send to Contract C the selected numbers. The Contract C verifies if number $s$ matches the sha3($s$) submitted in the first step meets the parameters, and if yes, then the Contract C will save into the seeds of the Function that will generate random numbers.

• Calculate the random number, return deposit and send rewards
After collecting all $s_i$, apply Function $f(s_1, s_2, ..., s_n)$ as the final random numbers, write them into the storage of Contract C, and return the result to all contracts that required this random sequence.

Integrating RANDAO system design into the smart contract, the operation procedure of the random number generation becomes Algorithm 3. In the algorithm 3, $key$ is the generation key for the final random number, written down by the players themselves. $key_t$ is the entire key set stored in RANDAO. $T$ is the valid time period for users to upload random number generation keys, and $t$ is the current timestamp. $f$ represents the random number generation function of the RANDAO system.

---
**Algorithm 3** RANDAO-Integrated RNG Process
---
**Require:** User account address $\epsilon$;
**Ensure:** A set of k random results ($R = r_1, r_2, ..., r_k$);
1: **while** $t \leq T$ **do**
2:     $\epsilon$ pays RANDAO $m$ ether as a deposit;
3:     $\epsilon$ uploads an arbitrary address to RANDAO;
4: **end while**
5: $\epsilon$ send a transaction with $key$ to RANDAO;
6: **if** $key$ in $key_t$ **then**
7:     $R = f(key_t)$;
8:     RANDAO returns deposit;
9: **else**
10:     $R = \emptyset$
11: **end if**
    **return** $R$;
---

It is essential to mention that $key_t$ does not only include the keys collected from players in this particular lottery event, but also keys collected from other events that are stored in the RANDAO database. However, using RANDAO services requires players to become members of RANDAO, and it is necessary to pay an amount of fees to generate random numbers each time, which disobeys the design priciple of DeLottery. Moreover, the centralized design of the RANDAO system is not acceptable in DeLottery, since RANDAO holds the deposit for a certain period of time during every generation period of random numbers. Similar trust problems may also lie behind. Thus, we would simply apply the algorithms of RANDAO into DeLottery, making it decentralized.

## 4. LOTTERY SYSTEM DESIGN

Our design tries to take both security and independence of a third party into consideration. The aim of DeLottery is to simplify the lottery event procedure to be as close to the core lottery procedure as possible, avoiding redundant steps. Here we show how DeLottery is designed by using smart contract in blockchain. The high phase system structure is shown in Figure 1. Players are connected in blockchain using their digital devices as nodes on the chain. And the lottery smart contract is deployed on the same blockchain. All players participate in the same lottery event, so they share one smart contract deployed on an arbitrary account. Then the players and the lottery system start interacting.

Consider that there are many users in a lottery system, this contract is deployed by an arbitrary user and every other user can participate in this contract interaction.

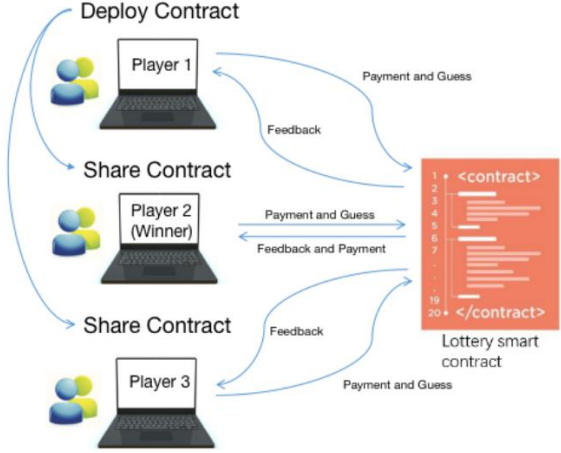

Figure 1: A Lottery System Diagram

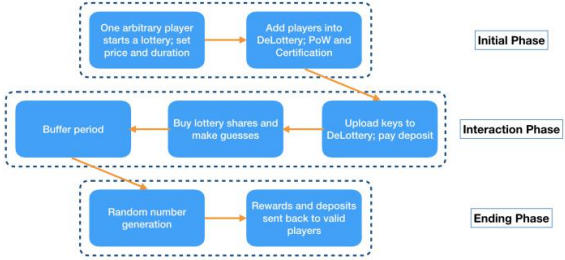

Figure 2: State Diagram

At the very beginning of the lottery procedure, the arbitrary host is the only account deployed with DeLottery. We design the lottery system using a high phase finite state machine diagram shown in the Figure 2. The seven steps in DeLottery procedure can be summarized in three phases with each phase possessing two steps: initial phase, interaction phase, and ending phase. Phases are indicated by different rows in Figure 2.

Based on the diagram and phases we designed, to store the critical data of the lottery system and users and to allow the users to interact with the system, the contract should have the following several variables as shown in Table 2. These variables are set to support the complete lottery process, and to ensure that the system is decentralized. As a typical lottery process, it contains all variables in Table 1, as well as specialized variables included in DeLottery.

## 4.1 Initial Phase

### 4.1.1 One arbitrary player starts DeLottery.

The very first step is to choose one player arbitrarily from all potential players who want to participate in the lottery event as the address the DeLottery contract. Also, the duration for the bet and the price of each share of lottery are decided in this step. This procedure is essential in order to keep the decentralized feature of DeLottery.

Table 2: DeLottery System Parameters

| Name | Type | Description |
|---|---|---|
| $n$ | $n \in \mathbb{N}$ | the number of players involved in the lottery process |
| $C$ | nonempty set | the set of all possible guesses of the lottery (treated as whole set) |
| $f_i$ | $f_i \in \mathbb{R}, i \in \{1, ..., n\}$ | the original amount of money of player $i$ |
| $g_i$ | $g_i \subset \mathbb{R}, i \in \{1, ..., n\}$ | the set of guesses made by player $i$ |
| $G$ | $G \subset C$ | the set of possible values of the winning results |
| $W$ | $W \subset G$ | the set of winning results of this lottery event |
| $t$ | $t \in \mathbb{R}$ | transaction fee deducted by the blockchain service provider |
| $r$ | $p \in \mathbb{R}$ | the total value of the accumulated reward |
| $s$ | $p \in \mathbb{N}$ | price of each guess |
| $P$ | nonempty set $P = \{p_i, i \in [1, n]\}$ | the total set of players who already deployed DeLottery |
| $p_i$ | $p_i \in P, i \in [1, n]$ | a player who deployed DeLottery |
| $A$ | $A \subset P$ | the set of activate players who certificate new players |
| $\tau$ | $\tau \in \mathbb{R}^+$ | the time since DeLottery is deployed |
| $d_i$ | $d_i \in \mathbb{R}^+$ | the deposit that player $i$ pay to DeLottery |
| $k$ | $k \in \mathbb{R}^+$ | the security factor of DeLottery |
| $\phi$ | $\phi \in \mathbb{R}^+$ | the total value of the accumulated ether captured from illegal accounts or requests |
| $B$ | $B \subset P$ | the set of all banned illegal player accounts |

### 4.1.2 Adding all players into the event.

Adding all players into the event. This aim of this step is to join all players in the same blockchain which DeLottery is deployed, as well as to maintain the safety of the system from potential Sybil attacks. Every time a new account wants to participate in the lottery event, a PoW is first done to guarantee that the account is real and capable for the lottery event, and identity certification is done between players to prevent Sybil attacks. We take the amount of certification work that needs to be done when the player group is large into consideration, and do optimization to the certification procedure. Algorithm 4 shows how all players are added. In the algorithm, $P = \{p_1, ..., p_n\}$ is the set of all players that are already added to DeLottery. A indicates the active certification accounts, such that $A \subset P$ and Card($A$) $\leq k$, $k \in \mathbb{N}$. Each $p_i$, $i \in [1, n]$ has a property *auth*, indicating whether player $i$ has authorized another player $j$ to certificate new members, and a property *time*, indicating the timestamp when player $j$ is authorized and entered the player group.

**Algorithm 4** Player Addition and Certification

**Require:** Potential player account address $\delta p$;
1: Do Proof of Work to $\delta p$;
2: **if** $\delta p$ pass PoW **then**
3:    **if** $\forall p_i \in A$ certification is successful **then**
4:       $P = P \bigcup \delta p$;
5:       **if** Card($A$) = $k$ **then**
6:          Find $p_o$ with $p_o.time \geq p_j time$, $j \neq o$;
7:          Replace $p_o$ with $\delta p$ in $A$;
8:       **else**
9:          $A = A \bigcup \delta p$;
10:       **end if**
11:    **else**
12:       Reject $\delta p$;
13:    **end if**
14: **else**
15:    Reject $\delta p$;
16: **end if**

## 4.2 Interaction Phase

### 4.2.1 Players upload generation keys to DeLottery.

Each player uploads a number between $-2^{63}$ to $2^{63}$ to DeLottery, and sends DeLottery $t$ ethers as deposit. The amount of $t$ is decided by equation (1).

$$t = \max(s \cdot 10^{\ln k}, \frac{f_i}{k}) \quad (1)$$

and $f_i$ is the total property of player $i$. $k$ is a the security factor of DeLottery, deciding how much each player should pay for the deposit. A typical value of $k \in (1, 2)$. Paying a large amount of deposit ensures the security of DeLottery from Sybil attacks, and restricts players from being too addicted to lottery. The procedure of key uploading is described with more details in Algorithm 5.

### 4.2.2 Players buy lottery shares.

For an arbitrary player who buys $m$ shares of lottery, the remaining ether in the account is calculated through equation (2).

$$f_i \leftarrow f_i - m \cdot (s + t) \quad (2)$$

The transaction fee is essential to be paid to the blockchain service provider, and the usual ratio of transaction fee is $10^{-12}$ the value of $s$. All fees are collected in the pool of DeLottery where the contract is deployed, and they are kept in frozen state. The prize pool of the DeLottery event is calculated by equation(3).

$$r = \phi + \sum_{i=1}^{n} d_i \quad (3)$$

### 4.2.3 Buffer period.

The fundamental purpose of buffer period is to deal with potential synchronization issues, and ensure that all guess activities are done before the lottery results are generated. After buffer period, no guess requested are allowed.

**Algorithm 5** Key uploading

**Require:** Player account address $p_i$;
1: $p_i$ generates $key \in \mathbf{R}$;
2: $p_i$ transfers $key$ to DeLottery;
3: $p_i p$ give deposit $d_i$ to the account where DeLottery is deposited;
4: DeLottery transfer the time remaining $time$ to $p_i$;
5: **while** current time $\leq time$ **do**
6:    $p_i$ send a transaction to DeLottery with $key$;
      **return** Success
7: **end while**
8: **if** $\tau > time$ **then**
9:    $\phi + = d_i$
      **return** Failure
10: **end if**

## 4.3 Ending Phase

### 4.3.1 Random number generation.

The random number generation returns the result that wins the lottery. The forms results and lottery shares adjust with different types of lotteries, but in general, the random number generation takes $n$ - Card($B$) key values into consideration, and generates the random result according to equation (4).

$$W = f(g_i, \forall i \text{ s.t. } p_i \in (P - B)) \quad (4)$$

### 4.3.2 Rewards are sent back to players.

In this step, for any arbitrary valid player $p_i$, if the player wins the lottery, then certain share of the accumulated reward will be sent to the player. The detailed procedures of this step is shown in Algorithm 6.

**Algorithm 6** Sending back rewards to a player

**Require:** A player account address $p_i$, total number of shares that wins the lottery $N_w$;
**Ensure:** Remaining ether of the player $f_i$;
1: **if** $g_i \bigcap W \neq \emptyset$ **then**
2:    $f_i$ += $\frac{\text{Card}(g_i \bigcap W) \cdot (\phi + \tau)}{N_w}$;
3: **else**
4:    $f_i$ remains unchanged;
5: **end if**
      **return** $f_i$

## 5. TRUST, FAIRNESS AND SECURITY

In this decentralized system, we can see several advantages over the traditional systems and centralized online systems:

Players have to pay only $10^{-12}$ times the total bet for the service of blockchain. Although this small amount of money is transferred to miners of blocks, but it seems ignorable comparing to the profit of lottery companies or casinos.

Comparing to traditional lottery systems, DeLottery is easier to deploy, convenient to use and efficient during transfer. Besides these factors, there are three major advantages that DeLottery has over previous lottery system designs: trust, fairness and security issues.

### 5.1.1 Trust

The rules and regulations are guaranteed to be followed automatically since they are programmed in to DeLottery. So every participant no longer needs to put their trust on any third parties.

### 5.1.2 Fairness

DeLottery no longer has a center. Although the system is started by some user, the user has exactly zero privilege after the initiation.

### 5.1.3 Security

DeLottery uses a RANDAO way to generate random numbers, which is a reliable and safe approach. This approach makes the system free from node attacks from transaction nodes and Sybil attacks, thus improving the overall stability of the lottery system.

## 6. CONCLUDING REMARKS

We design and implement DeLottery, a novel type of lottery systems based on blockchain and smart contract technologies. The lottery system has a decentralized feature, and is capable of ensuring user trust, fairness and security demands. We adopt blockchain technology and smart contract, as well as RANDAO [10] in the design of DeLottery. The whole system follows Initial Phase, Interaction Phase and Ending Phase, each including two detailed steps. The six steps are easy to implement by Solidity, and ensure efficiency, since each step has a time complexity no more than *O(n)*. This feature provides DeLottery with a wide range of applications, especially when dealing with huge amount of data.

In DeLottery, the smart contract replace the role of a third party. There is no need for a lottery company or a third party to supervise and lead this activity. On the other hand, DeLottery is secure in its design, and has adequate resistance to node attacks and Sybil attacks, which are two major attack methods to the blockchain system. The winner of the lottery is produced randomly and the prizes will surely be sent to the them. The fairness of lottery and gambling is guaranteed. The lottery system has become completely decentralized.

DeLottery is implemented in Solidity and deployed on Ethereum. Future research will be aimed at performance optimization, as well as model migration, so that the design of DeLottery can contribute more to its field.

## 7. ACKNOWLEDGEMENT

This work has been partially supported by NSFC Grant 61572323 and Grant-in-Aid for Scientific Research from Japan Society for Promotion of Science (JSPS) No. 26280027. Jie Li is the corresponding author.